\documentclass[fleqn,usenatbib]{mnras}

\usepackage{newtxtext,newtxmath}

\usepackage[T1]{fontenc}

\usepackage{graphicx}

\usepackage{multicol}	
\usepackage[english]{babel}
\usepackage{epstopdf}
\usepackage{ae,aecompl}
\usepackage{hyperref}
\usepackage{amsmath}    

\def\mpc{\,h^{-1}{\rm Mpc}}

\def\msun{\,h^{-1}{\rm M}_\odot}

\usepackage{color}

\makeatletter

\newcommand{\Rmnum}[1]{\expandafter\@slowromancap\romannumeral #1@}
\makeatother

\title[filaments in the cosmic web]{Statistical properties of filaments in the cosmic web}

\author[Youcai Zhang]{Youcai Zhang$^{1}$\thanks{yczhang@shao.ac.cn},
Hong Guo$^{1}$, Xiaohu Yang$^{2,3}$, Peng Wang$^{1}$
\\
$^{1}${Shanghai Astronomical Observatory; Nandan Road 80, Shanghai 200030,
  China} \\
$^{2}$Tsung-Dao Lee Institute and Key Laboratory for
Particle Physics, Astrophysics, and Cosmology, Ministry of Education, \\
    ~~Shanghai Jiao Tong University, Shanghai 200240, China \\
$^{3}$Department
of Astronomy, School of Physics and Astronomy and Shanghai Key Laboratory for Particle Physics and Cosmology, \\
~~Shanghai Jiao Tong University, Shanghai 200240, China
}

\begin{document}
\label{firstpage}
\pagerange{\pageref{firstpage}--\pageref{lastpage}}
\maketitle

\begin{abstract}
In the context of the cosmological and constrained ELUCID simulation, this study explores the statistical characteristics of filaments within the cosmic web, focussing on aspects such as the distribution of filament lengths and their radial density profiles. Using the classification of the cosmic web environment through the Hessian matrix of the density field, our primary focus is on how cosmic structures react to the two variables $R_{\rm s}$ and $\lambda_{\rm th}$. The findings show that the volume fractions of knots, filaments, sheets, and voids are highly influenced by the threshold parameter $\lambda_{\rm th}$, with only a slight influence from the smoothing length $R_{\rm s}$. The central axis of the cylindrical filament is pinpointed using the medial-axis thinning algorithm of the COWS method. It is observed that median filament lengths tend to increase as the smoothing lengths increase. Analysis of filament length functions at different values of $R_{\rm s}$ indicates a reduction in shorter filaments and an increase in longer filaments as $R_{\rm s}$ increases, peaking around $2.5R_{\rm s}$. The study also shows that the radial density profiles of filaments are markedly affected by the parameters $R_{\rm s}$ and $\lambda_{\rm th}$, showing a valley at approximately $2R_{\rm s}$, with increases in the threshold leading to higher amplitudes of the density profile. Moreover, shorter filaments tend to have denser profiles than their longer counterparts.

\end{abstract}

\begin{keywords}
large-scale structure of universe -- methods: statistical --
  cosmology: observations
\end{keywords}

\section{Introduction}\label{sec_intro}

On the order of Mpc scales, the distribution of galaxies and dark matter reveals a
striking gigantic network named {\tt cosmic web} \citep{Bond1996} consisting of knots
(or clusters), filaments, sheets (or walls) and voids, arising from the primordial
density fluctuations via gravitational instability \citep{Peebles1967,Zeldovich1970}. 
Filaments are highly intriguing features in the cosmic web and play an important role in matter transport as bridges between galaxies. The main challenges in
the study of the cosmic web are to precisely identify the filamentary structures and
quantitatively characterize the properties of filaments, such as the distribution of filament length and the profile of radial density \citep{Galarraga2020, Pfeifer2022, 
Zakharoval2023,Wang2024}.

In observation, cosmic filaments traced by galaxies are well drawn by galaxy redshift
surveys, such as the Center for Astrophysics redshift
survey \citep[CfA,][]{deLapp1986}, the Sloan Digital Sky
Survey \citep[SDSS,][]{Abazajian2009} and
the Dark Energy Spectroscopic Instrument \citep[DESI,][]{DESI2023}.  Furthermore,
cosmological simulations have been performed to present the evolution of dark matter,
which allows a more detailed study of filaments traced by the dark matter distribution
\citep{Springel2005Nature, Zhang2009, Dubois2014, WangZitong2024}.  Compared to
approximately
spherical structures, filaments are more complicated geometric structures associated
with a distinct direction in space, resulting in some challenges in identifying cosmic
filaments based on galaxies in observation or dark matter particles in simulation
\citep{Zakharoval2023}. 

Over recent decades, numerous methods have been formulated for the detection of cosmic filaments through geometric or topological evaluations of the discrete distributions of particles in simulations or observed galaxies \citep{Stoica2005, Sousbie2011a, Sousbie2011b, Chen2015, Bonnaire2020}, including the analysis of the Hessian matrix of density, tidal, or velocity shear tensors \citep{Colombi2000, Arag2007, Hahn2007, FR2009, Hoffman2012, Cautun2013, Wang2020}, and the application of machine learning approaches \citep{Aragon2019, Buncher2020, Suarez2021, Carron2022}.
A detailed comparison of the different filament finders is presented in
\citet{Libeskind2018}.  In general, these filament detection algorithms are
designed for different purposes in studies. Assessing the relative success of these methods is challenging because there is no universally accepted definition of filaments \citep{Alina2022, Inoue2022, ZhangYikun2022, Aragon2023}.

With the help of filament detection methods, several studies of cosmic filaments
have been carried out to investigate the effects of the environment on galaxy formation and
evolution \citep{Kuutma2017, Chen2017}. By applying the DisPerSE filament
identification algorithm \citep{Sousbie2011a} to galaxies at redshift $z \simeq 0.7$
in the VIPERS survey, \citet{Malavasi2017} found that more massive or passive
galaxies are closer to the filaments than less massive or active galaxies. 
Based on the filament catalogs traced by the DisPerSE algorithm, the lower specific
star formation rates with decreasing distance to the filaments are also confirmed in
the GAMA survey with $ 0.02\leq z \leq 0.25$ \citep{Kraljic2018}, in the COSMOS
with $0.5<z<0.9$ \citep{Laigle2018}, and in the SDSS with $z \sim 0.1$ \citep{Winkel2021}.

Recent studies have focused on the statistical properties of cosmic filaments, such as their lengths and radial density profiles \citep{Malavasi2020, YangTianyi2022, Galarraga2020, Galarraga2023}. Observationally, using filament samples identified by the Smoothed Hessian Major Axis Filament Finder (SHMAFF) for SDSS ($z \sim 0.1$) and DEEP2 ($z \sim 0.8$), \citet{Bond2010} and \citet{Choi2010} discovered that the distribution of filament lengths is roughly exponential and that the filament widths are heavily influenced by the smoothing scales used in the creation of the smoothed density. \citet{Wang2024} propose that filament radii should be identified at points where galaxy number density profiles have the minimal gradient around filaments. This approach provides a physical delineation of filament boundaries akin to the splashback radius observed in dark matter halos \citep{Diemer2014}. Using filaments traced by the DisPerSE algorithm for SDSS MGS and LOWZ+CMASS samples, \citet{Malavasi2020} examined how filament properties are affected by the selection of parameters in the DisPerSE algorithm. They determined that raising the persistence threshold can remove a substantial portion of short filaments and that the skeleton-smoothing process does not significantly alter the filament length distributions.

In simulation, based on filaments traced by DisPerSE from the large hydrodynamic
Illustris-TNG simulation, \citet{Galarraga2020} investigated the population of
filament lengths and radial density profiles. They claimed that short and long
filaments do not have the same radial density profiles. For the total population of
filaments from the MillenniumTNG simulations, \citet{Galarraga2023} found very
little evolution of the filament length distributions or radial density profiles. 
Using filaments connecting group-mass halo pairs in a dark matter only simulation,
\citet{YangTianyi2022} provided a universal function with four parameters to
fit the filament radial density profiles. They found that the scale radii
parameters have little evolution with redshift but increase with filament
length, indicating that filaments of different lengths actually follow different
evolution histories.

However, the statistical properties of filaments depend not only on the
identification algorithms but also on the different tracers, such as dark-matter particles or galaxies. Unlike relatively spherical haloes, filaments are more complex because of their elongated shape. The measurement of filament lengths is influenced by the way their endpoints are determined, a process often impacted by the settings used in their identification methods \citep{Bond2010, Malavasi2020}. \citet{Rost2020} compared three SDSS filament catalogues
identified by different identification techniques \citep{Tempel2014, Martinez2016, 
Pereyra2020}. It was observed that the characteristics of filaments vary significantly with the choice of identification methods. Using the DisPerSE algorithm on the semi-analytical model's results, \citet{Zakharoval2023} explored the attributes of filaments derived from both model galaxies and dark-matter particles. The investigation revealed that filaments associated with dark-matter particles consistently exhibit greater lengths compared to those derived from galaxies. The process of extracting filaments was asserted to be heavily dependent on the tracers used and the settings implemented in the DisPerSE algorithm. 

The statistical characteristics of filaments, as observed and simulated, are still not well understood. This study examines these characteristics through the use of the COWS algorithm \citep{Pfeifer2022} and the ELUCID simulation \citep{WangH2014,WangH2016, WangH2018}, which are constrained by the density field of the galaxy distribution, thereby connecting the simulated dark matter distribution with the actual observations of the galaxy distributions \citep{Zhang2021a,Zhang2021b,Zhang2022}, with a specific focus on offering a  quantitative analysis of the statistical properties of filaments. The COWS method applies a medial axis thinning algorithm \citep{Zhang1984, Lee1994} to the cosmic web's velocity shear tensor (V-web) to identify the filament skeleton, maintaining topological and geometric constraints. In this study, we use the Hessian matrix of the density field to classify the cosmic web environment instead of using the velocity shear tensor data set used in \citet{Pfeifer2022}.  Additionally, the connection between the ELUCID simulation and SDSS observations \citep{YangX2018} allows further investigation into differences in filament extraction and analysis across both dark-matter and galaxy distributions in future studies.

The structure of this paper is organised as follows. In Section~\ref{sec_data},
we provide an overview of the ELUCID $N$ body simulation, the cosmic web
classification algorithm based on the smoothed density field, and the COWS
filament finding method. In Section~\ref{sec_result}, we present a statistical
study of the filamentary structures of the cosmic web, such as the filament
length distribution and the radial density profile. Finally, we summarise our results
in Section~\ref{sec_summary}. 

\section{Data and Method}\label{sec_data}

The data used in this paper are from the constrained ELUCID simulation \citep{WangH2014, WangH2016, WangH2018}, which can reproduce the
density field of the nearby universe generated from the galaxy group distributions
in the SDSS observation \citep{YangX2007, YangX2012}. As expected, the statistical properties of cosmic
filaments are well reproduced in the ELUCID simulation compared to the SDSS
observation \citep{WangH2016}.

The ELUCID simulation is a dark matter only cosmological simulation with $3072^3$
particles evolved in a periodic box of $500 \mpc$ on a side from an initial
redshift of $z=100$ to the present. The initial condition of the ELUCID simulation
is constrained by the density field of the galaxy distribution of Sloan Digital
Sky Survey Data Release 7 \citep[SDSS DR7;][]{Abazajian2009}, employing the Hamiltonian Markov Chain Monte Carlo (HMC) algorithm combined with
Particle Mesh (PM) dynamics \citep{WangH2014}. The initial density field
evolved to the present day using L-Gadget, a memory-optimised version of Gadget-2 
\citep{Springel2005}, with $100$ snapshot output from $z=18.4$ to $z=0$. In this
work, we only use the distribution of dark matter particles at redshift $z=0$. 
The mass of each dark-matter particle is $3.1 \times 10^8 \msun$. The adopted
cosmological parameters are $\Omega_m = 0.258$, $\Omega_b = 0.044$, $\Omega_\Lambda = 0.742$, $h = 0.72$, $n_s = 0.963$, and $\sigma_8 = 0.796$.

 In the simulation, the standard Friends-of-Friends (FOF) method is applied to the particle data to generate FOF haloes, using a linking length of $b = 0.2$ times the average particle separation. Subsequently, the SUBFIND algorithm splits each FOF halo into individual subhaloes by applying a gravitational binding procedure during the unbinding phase. The total count of subhaloes in the ELUCID simulation is $48,846,170$ with masses exceeding $10^{9.8} \msun$.

The observational data used in this research comes from SDSS \citep{Abazajian2009}, known as one of the most important astronomical surveys. Utilizing multi-band imaging and spectroscopic data from SDSS DR7, \citet{Blanton2005} created the New York University Value-Added Galaxy Catalogue (NYU-VAGC) incorporating a unique set of enhanced reductions. From the NYU-VAGC, we gathered a total of $639,359$ galaxies with redshifts in the interval $0.01 \leq z \leq 0.2$, redshift completeness ${\cal C}_z > 0.7$, and extinction-corrected apparent magnitudes brighter than $r \leq 17.72$. The coverage area of $639,359$ galaxies spans $7,748$ square degrees, divided into two sections: an extensive continuous area in the Northern Galactic Cap and a smaller section in the Southern Galactic Cap. The galaxy distribution in SDSS DR7 is linked to the dark matter structure pinpointed by the constrained ELUCID simulation \citep{YangX2018}.

\subsection{Cosmic web classification algorithm}\label{sec_cw}

\begin{figure}
\includegraphics[width=0.5\textwidth]{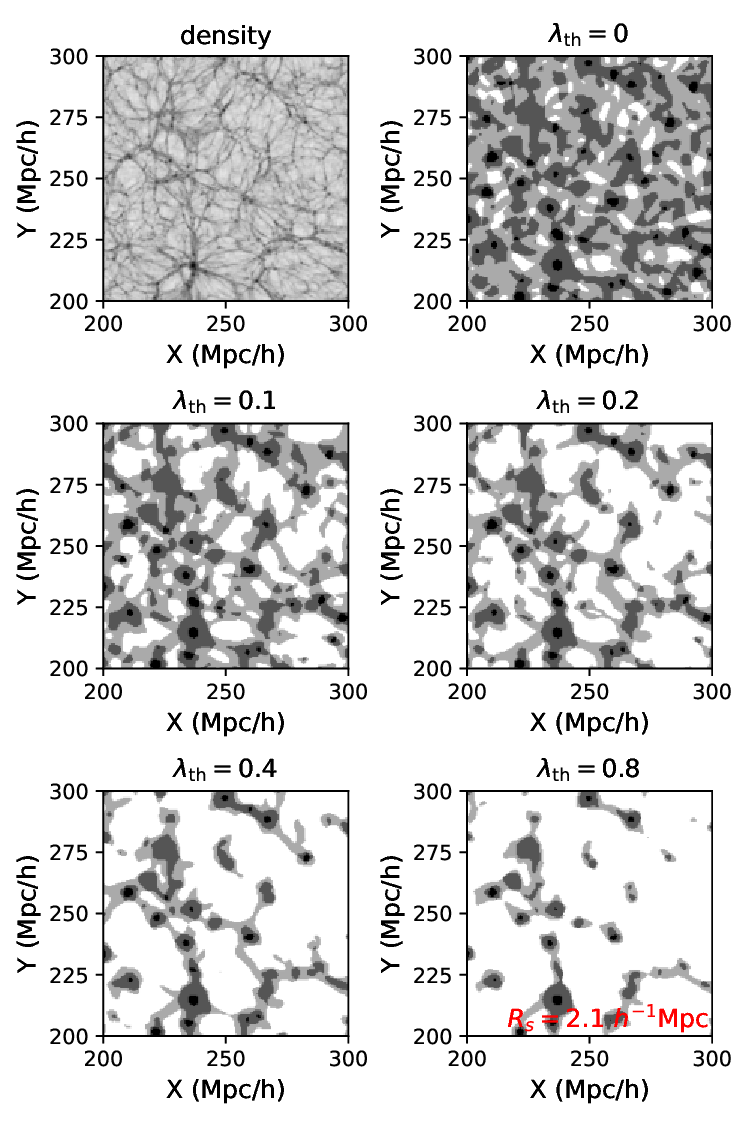}
\caption{Density distribution and the cosmic web classifications in a slice of
width $0.5 \mpc$. The upper-left panel shows the projected logarithmic density 
distribution, where darker areas indicate higher density. The other panels show
the environment classifications with the smoothing length $R_{\rm s} = 2.1\mpc$ 
in different values of the thresholds $\lambda_{\rm th} = 0, 0.1, 0.2, 0.4, 0.8$,
according to which the grid cells are classified into knots (black), filaments 
(dark grey), sheets (light grey), and voids (white), respectively.}
\label{fig:grid_cw}
\end{figure}

\begin{figure}
\includegraphics[width=0.5\textwidth]{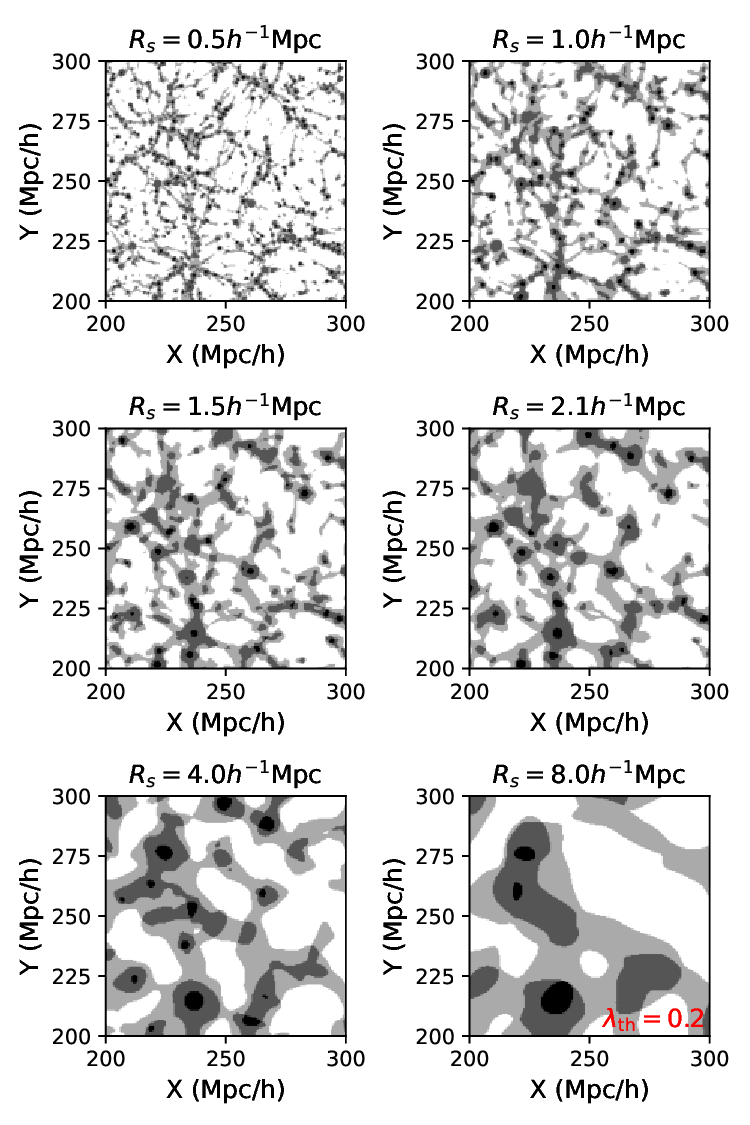}
\caption{Similar to Figure~\ref{fig:grid_cw} but for different Gaussian smoothing
lengths $R_{\rm s} = 0.5,1.0,1.5,2.1,4.0,8.0\mpc$ at fixed threshold 
$\lambda_{\rm th}=0.2$.
}
\label{fig:grid_smooth}
\end{figure}

The cosmic web type at any point in space can be classified by the Hessian
matrix of the density field (hereafter D-web), which was pioneered by
\citet{Colombi2000}. The Hessian matrix, adjusted by a negative normalization, of the smoothed density field is specified for each grid cell as follows:
\begin{equation}\label{eqn:hessian}
H_{ij} = - \frac{R_{\rm s}^2}{\rho_{\rm mean}} \frac {\partial^2 \rho_{\rm s}}
{\partial x_i \partial x_j }, 
\end{equation}
where $\rho_{\rm s}$ represents the smoothed density field employing a Gaussian filter with a smoothing length $R_{\rm s}$, and $\rho_{\rm mean}$ denotes the universal average matter density. A negative sign is incorporated on the right-hand side of Equation~\ref{eqn:hessian} to ensure that positive (negative) eigenvalues indicate collapsing (expanding) matter, aligned with the eigenvalue signs of the tidal tensor in the dynamic model \citep{Hahn2007}. 
The classification of environments is determined by the count of eigenvalues that exceed a specified threshold $\lambda_{\rm th}$. Specifically, the counts of $0, 1, 2,$ or $3$ are indicative of voids, sheets, filaments, and knots, in the order \citep{FR2009}.

To calculate the Hessian matrix of the density field $H_{ij}$, Cloud-in-Cell (CIC) interpolation
of $1024^3$ cells is used to construct the density field
$\rho_{\rm cic}$ for $3072^3$ particles in a periodic box of $500\mpc$, 
resulting in an average of $27$ particles in each cell to ensure that few if any cells
are empty. The CIC density field $\rho_{\rm cic}$ is then smoothed with a Gaussian kernel with width $R_{\rm s}$ to generate the smoothed density $\rho_{\rm s}$, to calculate the eigenvalues of the Hessian symmetric matrix
in Equation~\ref{eqn:hessian} at the position of each cell. Following convention, eigenvalues are denoted by $\lambda_1, \lambda_2$ and $\lambda_3$
($\lambda_1>\lambda_2>\lambda_3$), corresponding to the eigenvectors
$\hat{\boldsymbol{e}}_1, \hat{\boldsymbol{e}}_2$ and $\hat{\boldsymbol{e}}_3$. 
The cosmic web type for each cell is classified by counting the number of
eigenvalues with $\lambda_i > \lambda_{\rm th}$, where $i=1,2,$ or $3$.
Note that the environment classification is local for each cell, and the
collection of neighboring cells with the same cosmic web type defines the
geometrical constructions named voids, sheets, filaments, or knots. 
For filament cells, the eigenvectors $\hat{\boldsymbol{e}}_3$ can be used
to indicate the directions of the filaments \citep{Zhang2009, Zhang2013, Zhang2015}. 

Obviously, the D-web method depends on two free parameters $R_{\rm s}$ and
$\lambda_{\rm th}$, which are typically chosen such that the environment
classification reproduces the visual impression of the cosmic web. 
Figure~\ref{fig:grid_cw} shows the density distribution and the classifications of
the cosmic web at fixed smoothing length $R_{\rm s} = 2.1\mpc$ for different
thresholds $\lambda_{\rm th} = 0, 0.1, 0.2, 0.4$, and $0.8$, where the grid
cells are classified into knots (black), filaments (dark grey), sheets (light grey) and voids (white), respectively. As shown in Figure~\ref{fig:grid_cw}, for $\lambda_{\rm th} = 0$, the volume filling fraction of the voids is very small, which contrasts with the visual impression of the voids in the density
distribution. Obviously, the volume fraction of the voids increases with the
threshold. For $\lambda_{\rm th} \geq 0.1$, the voids occupy a large fraction
of the simulated volume, which appears to reproduce the visual impression of the
cosmic web in the simulation.

For cells in a grid, the classifications of the cosmic web depend not only on
the threshold parameter $\lambda_{\rm th}$, but also on the smoothing length
$R_{\rm s}$. Figure~\ref{fig:grid_smooth} shows the cosmic web classifications
for different smoothing lengths $R_{\rm s} = 0.5,1.0,1.5,2.1,4.0,8.0\mpc$ at a
fixed threshold $\lambda_{\rm th}=0.2$. Figure~\ref{fig:grid_smooth} shows
that increasing the smoothing lengths eliminates a large fraction of less
significant filaments. The smoothing procedure can remove sharp directional changes
and nonphysical
edges in filaments due to the noise of the density distribution. In this study, 
the cell size of the grid is approximately $0.5\mpc$. 
Using larger smoothing lengths can alleviate the effect
of shot noise on the filament geometry.
In the following analysis, we focus mainly on the cosmic
web classifications identified with the smoothing length $R_{\rm s} \geq 1 \mpc$
and the threshold parameters $\lambda_{\rm th} \geq 0.1$.

\subsection{Filament finding algorithm: COWS}\label{sec_cows}

\begin{figure}
\includegraphics[width=0.5\textwidth] {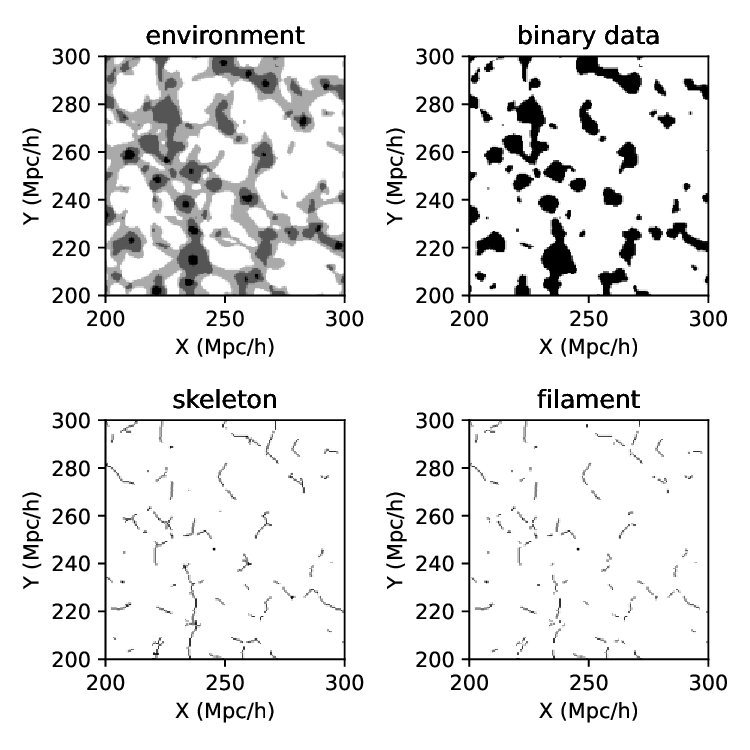}
\caption{Illustration of the COWS method to find the filament spine based on
the cosmic web classifications identified by the D-web method. Upper-left: 
Cosmic web environment identified by the Hessian matrix of the 
density field. Upper-right: Binary data with a value of $1$ (black region) 
identified as knots and filaments, and a value of $0$ (white region) identified 
as sheets and voids. Lower-left: Skeleton classification using the medial axis 
thinning method. Lower-right: Filaments identified by removing cavities (blobs) 
and junctions in the skeleton. }
\label{fig:slice_cow}
\end{figure}

\begin{figure}
\includegraphics[width=0.5\textwidth]{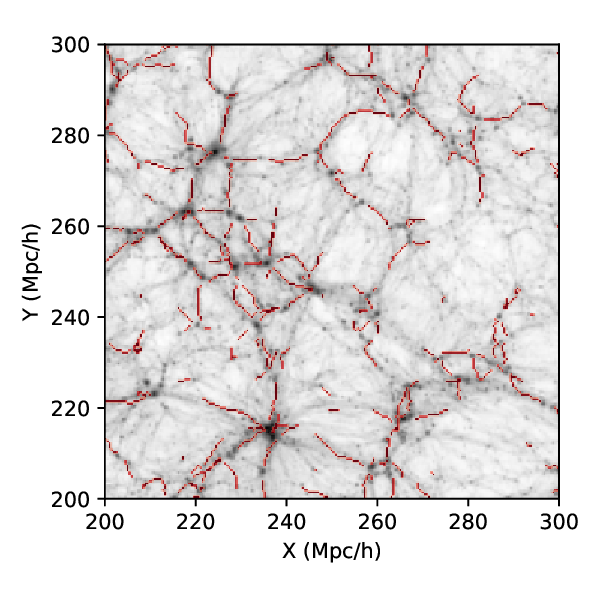}
\caption{Density and filament distribution in a slice of width $10 \mpc$ from
the ELUCID simulation. The filament spine in red is identified by the COWS 
method, based on the cosmic web classifications with $R_{\rm s} = 2.1 \mpc$ 
and $\lambda_{\rm th} = 0.2$.}
\label{fig:slice_filament}
\end{figure}

In this work, the filament axes are identified by a medial axis
thinning algorithm \citep{Lee1994} applied in the COWS
method\footnote{\href{https://github.com/SimonPfeifer/cows}
{https://github.com/SimonPfeifer/cows}} \citep{Pfeifer2022}. 

Figure~\ref{fig:slice_cow} presents a depiction of the COWS technique to identify filament spines within the ELUCID simulation. The upper left panel of Figure~\ref{fig:slice_cow} displays the input data in which the cosmic web classifications are determined using the Hessian matrix of the density field, as described in Section~\ref{sec_cw}, rather than employing the velocity shear tensor mentioned in \citet{Pfeifer2022}. As shown in the upper right panel of
Figure~\ref{fig:slice_cow}, the cosmic web classifications are then transformed
into binary data with a value of $1$ (black region) for knot and filament cells
and a value of $0$ (white region) for sheets and void cells. Next, the medial
axis thinning algorithm \citep{Lee1994}  is used to reduce large amounts of
binary data to representations that span one cell by identifying and removing
border cells while preserving the geometric and topological features of the
data. The lower left panel of Figure~\ref{fig:slice_cow} shows the medial axes
of the set of cells within the black region of the upper right panel. The medial
axes are often called the topological skeleton, which has the same connectivity
as the original data. Finally, individual filaments are defined by removing
any unwanted contaminants or features, such as junctions and hollow cavities,
from the topological skeleton.

Figure~\ref{fig:slice_filament} illustrates the distribution of dark matter density and filaments across a $10 \mpc$ slice of the ELUCID simulation, highlighting areas of greater density with darker shades. Generally, the resulting filaments in
red trace the majority of visible filamentary structures very well
in Figure~\ref{fig:slice_filament}. 

The axis thinning algorithm to identify filaments from the D-web has no
free parameter. However, the D-web method itself has two parameters, the
smoothing length $R_{\rm s}$ and the threshold $\lambda_{\rm th}$, which can
affect the filament catalogue extracted from the density distribution of dark
matter particles. It is interesting to investigate the dependence of the
statistical properties of filaments on the free parameters. For the case of
$R_{\rm s} = 2.1 \mpc$  and $\lambda_{\rm th} = 0.2$, the identified filaments are
coloured red as shown in Figure~\ref{fig:slice_filament}. Note that each
filament identified by COWS has two end points connected by a single path with
one cell width, which can be used to calculate the filament length and the
density profile perpendicular to the filament axis. 

\section{Results}\label{sec_result}
Using the filament catalogue identified by the COWS method from the ELUCID simulation, we focus mainly on the statistical properties of filaments, such as the filament length distributions and the density profiles.

\subsection{Volume fraction}
\begin{figure}
\includegraphics[width=0.5\textwidth]{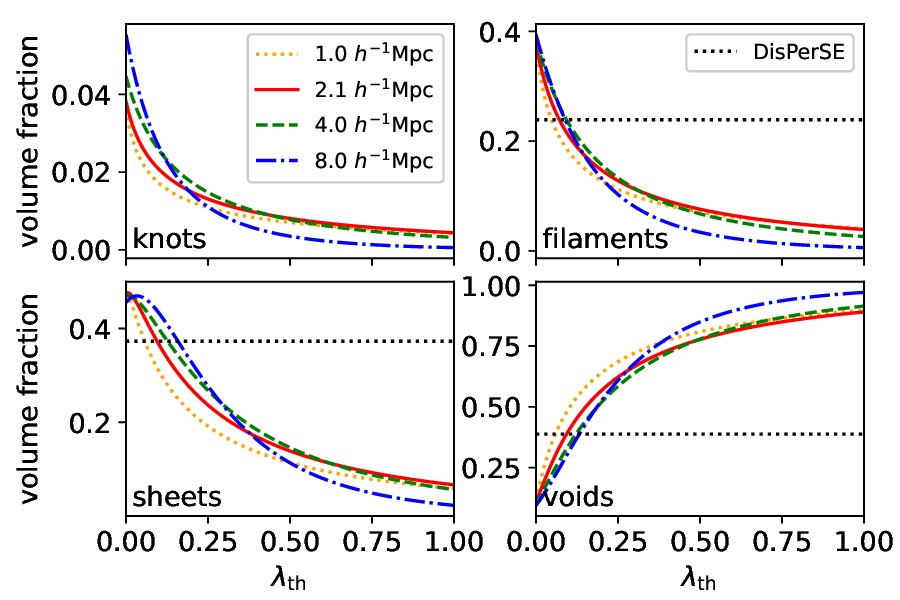}
\caption{Volume fractions of knots, filaments, sheets and voids as a function of
the threshold $\lambda_{\rm th}$ with different smoothing lengths $R_{\rm s} = 
1.0, 2.1, 4.0, 8.0 \mpc$. For comparison,
the black dotted line represents the volume fractions derived using the DisPerSE method, 
as detailed in Table 2 of \citet{Libeskind2018}.}
\label{fig:vff}
\end{figure}

Based on the cosmic web classification using the D-web method, we investigate the
dependence of the volume fractions for different types of cosmic webs on the free
parameters $R_{\rm s}$ and $\lambda_{\rm th}$. For a more detailed comparison of
different methods, we refer the reader to Table 2 and Figure 5 in \cite{Libeskind2018}.

Volume fractions are measured for knots, filaments, sheets, and voids by
varying the threshold $\lambda_{\rm th}$ from $0$ to $1$ at fixed smoothing
lengths $R_{\rm s} = 1.0, 2.1, 4.0, 8.0 \mpc$, as shown in Figure~\ref{fig:vff}. 
Generally, the volume fraction for each type of cosmic web is strongly dependent
on the threshold parameter $\lambda_{\rm th}$. For knots, filaments, and sheets,
the volume fractions decrease with increasing threshold $\lambda_{\rm th}$, 
while the volume fraction of the voids increases with increasing $\lambda_{\rm th}$.
At fixed $R_{\rm s}=2.1 \mpc$ represented by the solid red lines in
Figure~\ref{fig:vff}, the volume filling fraction of the filaments
decreases from $37.5\%$ to $3.9\%$ for increasing $\lambda_{\rm th}$ from
$0$ to $1$, while the volume fraction of the voids increases from $11.0\%$ to
$89.0\%$ for $\lambda_{\rm th}$ between $0$ and $1$.

Additionally, the volume fraction of each environment is weakly dependent on
the smoothing length. For the case of $\lambda_{\rm th}=0$, there is almost
no dependence on the smoothing length $R_{\rm s}$ for the volume fractions
of filaments, sheets, and voids. For the case of $\lambda_{\rm th} \gtrsim 0.4$,
the volume fraction of the filaments decreases with increasing smoothing length,
contrary to the trend of $\lambda_{\rm th} \lesssim 0.1$ where larger
smoothing lengths correspond to larger volume fractions, as shown in the upper right panel of Figure~\ref{fig:vff}.

\subsection{Filament orientation}
\begin{figure*}
\includegraphics[width=0.49\textwidth]{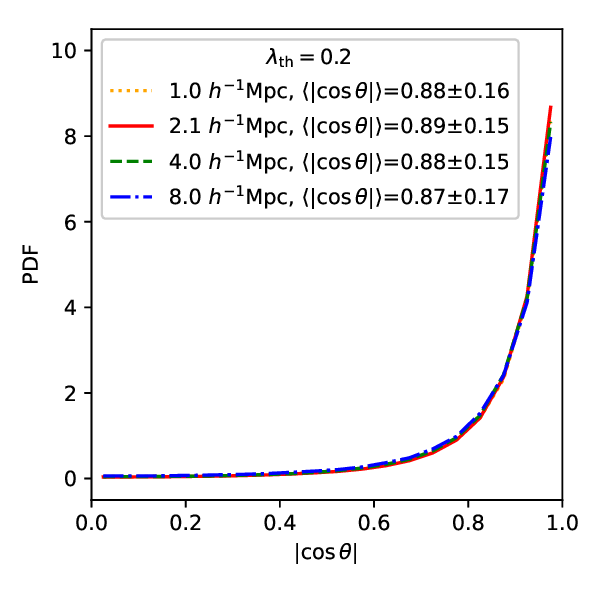}
\includegraphics[width=0.49\textwidth]{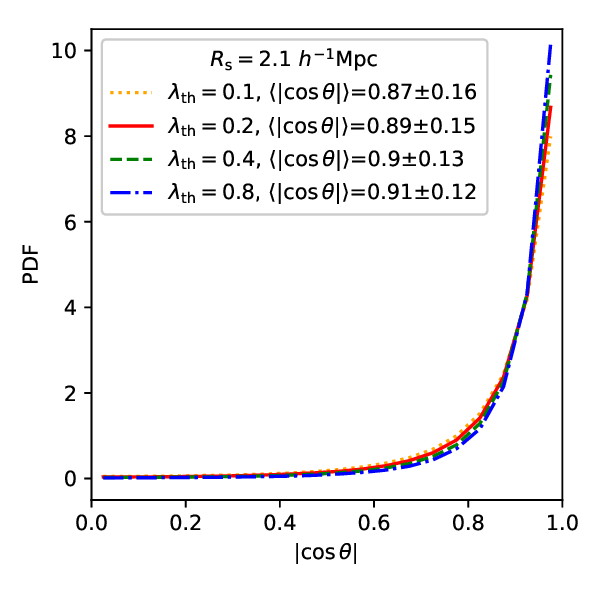}
\caption{Probability distribution function (PDF) of the cosine of the angle $\theta$
between the filament direction in COWS and the eigenvector 
$\hat{\boldsymbol{e}}_3$ in D-web. The left panel shows the PDF for different
$R_{\rm s}$ at fixed threshold $\lambda_{\rm th} = 0.2$, while the right panel
shows the PDF for different thresholds $\lambda_{\rm th}$ at fixed $R_{\rm s} 
= 2.1 \mpc$. The mean values of $|\cos \theta|$ and the standard errors
are indicated in the panels.}
\label{fig:alignment}
\end{figure*}

In this study, the filament axis is a single path with a cell width,
indicating the medial axis of the filament cells identified by
the D-web method. The filament axis in the COWS method is expected to be
aligned with the eigenvector $\hat{\boldsymbol{e}}_3$ of the Hessian density matrix in equation~\ref{eqn:hessian}. In the following analysis,
we investigate the directions of the filaments in the COWS method with
respect to the eigenvector $\hat{\boldsymbol{e}}_3$ in the D-web method.

In the COWS approach, a filament consists of two terminal points linked by a continuous chain of connected cells. Following \citet{Pfeifer2022}, the direction 
$\boldsymbol {d}_i$ of the filament 
cell $i$ is calculated by its two neighbours named $i+1$ and $i-1$,
\begin{equation}\label{eqn:direction}
\boldsymbol {d}_i = \boldsymbol{p}_{i+1}- \boldsymbol{p}_{i-1},
\end{equation}
where $\boldsymbol {p}_{i+1}$ and $\boldsymbol {p}_{i+1}$ are the positions 
of the two neighbour cells $i+1$ and $i-1$. For the terminal point in a filament, 
the direction $\boldsymbol {d}_i$ is set to be the vector between the end
point and its neighbour. 

In the D-web method, the eigenvalues of the density Hessian matrix are
calculated to quantify the collapsed or expanding matter. The corresponding
eigenvectors indicate the local orientation of the morphological structure. 
The filament cells in the D-web method are identified where the eigenvalues
are $\lambda_1 > \lambda_{\rm  th}$, $\lambda_2 > \lambda_{\rm th}$, and $\lambda_3 < \lambda_{\rm th}$.
The eigenvector $\hat{\boldsymbol{e}}_3$ of the density Hessian matrix can be used
to indicate the direction of the filament in the D-web. To verify the validity of
filament finding methods,
we calculate the alignment between the direction of the filament $\boldsymbol {d}_i$
and $\hat{\boldsymbol{e}}_3$. 

According to the definition of the filament direction given in Equation~\ref{eqn:direction}, for a filament cell with two neighboring cells, the connection between these two neighbors should be considered as an axis rather than a vector, because the order in which the neighbors are chosen does not matter. Therefore, the alignment is assessed by calculating the absolute value of the dot product between the filament direction and $\hat{\boldsymbol{e}}_3$. 
Figure~\ref{fig:alignment} shows the probability
distribution function (PDF) of the absolute cosine of the angle $\theta$ between
the direction of the filament $\boldsymbol {d}_i$ in COWS and the eigenvector
$\hat{\boldsymbol{e}}_3$ in D-web, where $\cos\theta = \boldsymbol {d}_i \cdot 
\hat{\boldsymbol{e}}_3/ | {\boldsymbol {d}_i} |$. In the absence of any
alignment, the probability distribution is flat with a constant value of $1$.
The average values of $|\cos\theta|$ are also shown in the panels of
Figure~\ref{fig:alignment}. The values of $\langle |\cos\theta| \rangle = 1$
imply that the direction of the filament $\boldsymbol {d}_i$ in COWS is parallel
to the eigenvector $\hat{\boldsymbol{e}}_3$ in the D-web, while the values of $\langle
|\cos\theta| \rangle = 0$ indicate perpendicular orientations.

The left panel of Figure~\ref{fig:alignment} shows the PDF for different
smoothing scales $R_{\rm s}$ with mean values and standard errors of $|\cos\theta|$. 
The filament orientations for each smoothing scale $R_s$ are compared to the corresponding eigenvectors
$\hat{\boldsymbol{e}}_3$ for those values of $R_s$. The alignment signals for different values of $R_s$ 
are nearly the same and remain largely unchanged with variations in smoothing lengths. The left panel of Figure~\ref{fig:alignment} illustrates that the orientations of the filaments are primarily aligned with the eigenvector $\hat{\boldsymbol{e}}_3$ in the D-web, validating the efficiency of the axis thinning algorithm in detecting the filament axis.

The right panel of Figure~\ref{fig:alignment} shows the PDF for different
thresholds $\lambda_{\rm th}$ at fixed $R_{\rm s} = 2.1 \mpc$. The alignment
signals become slightly stronger for increasing thresholds, with $\langle |\cos\theta|
\rangle = 0.91 \pm 0.15$ for $\lambda_{\rm th}=0.8$. Obviously, the filament
spines are strongly aligned with the eigenvector $\hat{\boldsymbol{e}}_3$ 
characterised by the underlying density field, indicating the validity of the COWS
methods.

\subsection{Filament length distribution}

\begin{figure*}
\includegraphics[width=0.49\textwidth]{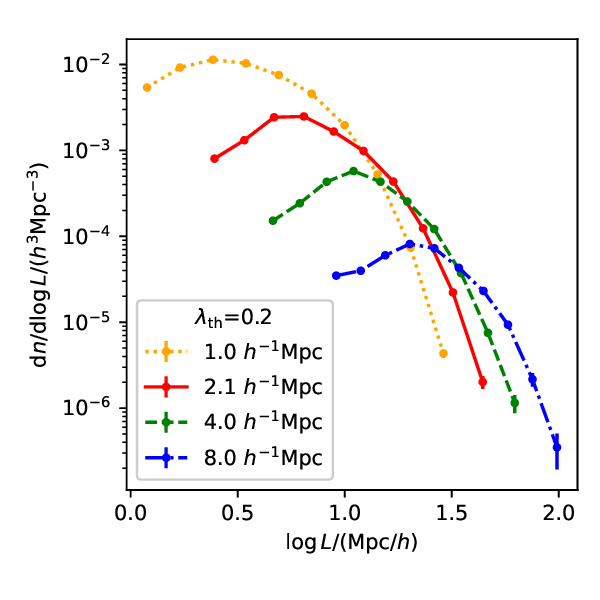}
\includegraphics[width=0.49\textwidth]{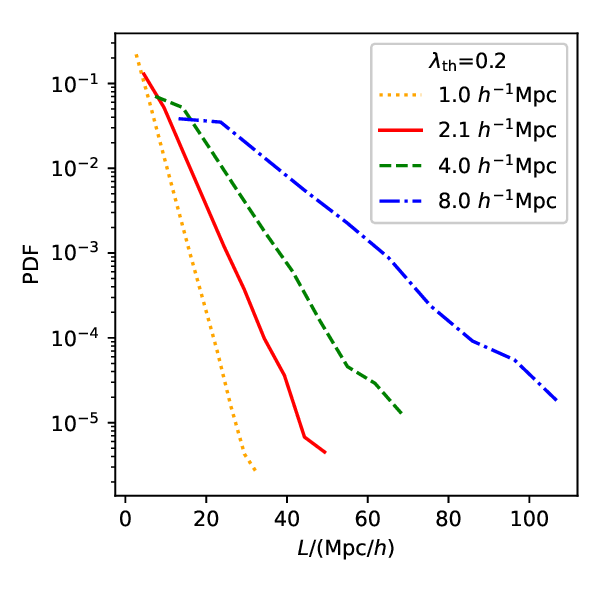}
\includegraphics[width=0.49\textwidth]{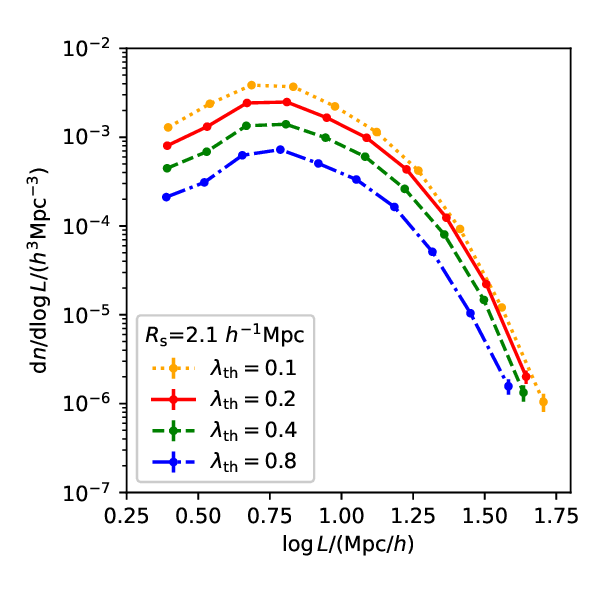}
\includegraphics[width=0.49\textwidth]{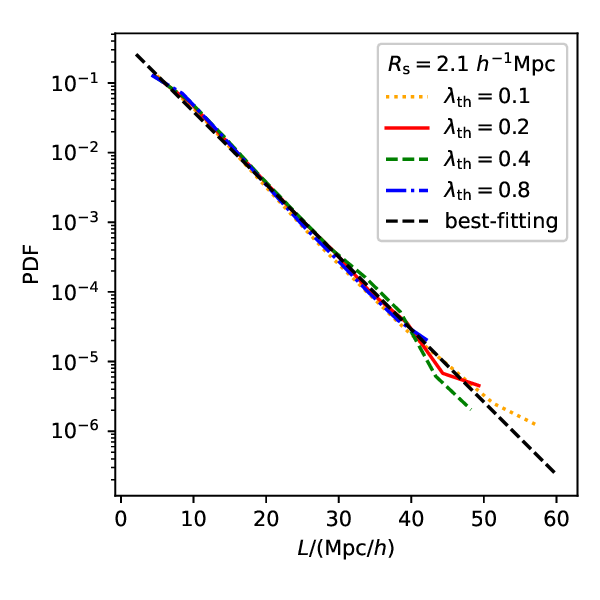}
\caption{Length distributions of filaments for varying the smoothing 
length $R_{\rm s}$ (upper panels) and the threshold $\lambda_{\rm th}$
(lower panels) in the simulation of $500\mpc$ on a side. The filament
length function is shown in the left panels with Poisson errors included.
The right panel shows
the probability distribution function of the filament lengths. In the lower right
panel, the black dashed line represents the best-fitting relation of the
form $\log f(L) = -0.1 L -0.37$, where $f(L)$ is the probability density
function.}
\label{fig:length}
\end{figure*}

In this section, we explore the distribution of lengths for filaments traced by dark matter particles in the ELUCID simulation. The length of the filament $L$ is defined as the distance along the filament path, which is delineated by the medial axis of the filament cells identified using the D-web method. To calculate the length of each filament, we sum the distances between the midpoints of adjacent cells and include an additional cell size to account for the two half-ends of the filament \citep{Pfeifer2022}.

An isolated spherical over-density should not be classified as a real filament, 
but the filament finder would identify an isolated spherical over-dense structure
as a filament of length on the order of the smoothing scale \citep{Bond2010, 
Choi2010}. Based on the filament catalogue, we have calculated the number
distribution of filament lengths for different smoothing scales $R_{\rm s}$.
We find that the number distribution of filament lengths exhibits a dramatic
drop-off below the smoothing length, in agreement with the results of
\citet{Bond2010}. Therefore, we hereafter exclude non-physical filaments whose lengths are shorter than the smoothing length $R_{\rm s}$. We focus only on
filaments of length $L \geq R_{\rm s}$, resulting in a total number of filaments
$N_{\rm f} = 979~552, 178~417, 35~353$ and $5~248$ for $R_{\rm s} = 1.0, 2.1, 
4.0$, and $8.0$, respectively. Obviously, the total number of filaments
decreases dramatically with increasing smoothing scales $R_{\rm s}$. 

The halo-mass distribution is conventionally described by halo mass function.
In spirit of the halo mass function, the filament length function (LF)  can be defined as
the number density of filaments per logarithmic unit of length. For comparison,
we also calculate the probability distribution function (PDF) $f(L)$ of filament lengths.
Figure~\ref{fig:length} shows the filament LFs and PDFs for varying the
smoothing length $R_{\rm s}$ (top) and the threshold $\lambda_{\rm th}$
(bottom) in the simulation of $500\mpc$ on one side.

The upper left panel of Figure~\ref{fig:length} shows the filament LF
for different smoothing scales $R_{\rm s} = 1.0, 2.1, 4.0$ and $8.0 \mpc$ at a
fixed threshold $\lambda_{\rm th}=0.2$. Generally, the median lengths of the filament increase with the
smoothing scales $R_{\rm s}$, which is evident in the ranges covered by
different filament length functions. Moreover, our observations indicate that LF typically reaches a maximum at $L \approx 2.5 R_{\rm s}$, for example, from approximately $5.3 \mpc$ at $R_{\rm s} = 2.1 \mpc$ to approximately $20.0 \mpc$ at $R_{\rm s} = 8.0 \mpc$. Furthermore, the number of short filaments decreases with
increasing smoothing scales, while the number of long filaments increases with
larger $R_{\rm s}$. 

The upper right panel of Figure~\ref{fig:length} shows the probability distribution
function of the filament lengths. Obviously, short filaments dominate
the catalogue of $R_{\rm s} = 1.0 \mpc$, while long filaments are easily detected in
the case of $R_{\rm s} = 8.0 \mpc$. Interestingly, we find that the logarithmic PDFs
of different smoothing scales are approximately proportional to the filament lengths
expressed by $\log f(L) \propto L$, where $f(L)$ is the PDF. 

Next, we investigate the dependence of the filament LF and PDF on the threshold
parameter $\lambda_{\rm th}$. The lower panels of Figure~\ref{fig:length} show the
filament LF and PDF for different thresholds $\lambda_{\rm th} = 0.1,
0.2, 0.4$, and $0.8$ at fixed $R_{\rm s} = 2.1 \mpc$. In the entire range, the
filament length functions decrease with increasing thresholds from $0.1$ to $0.8$, 
as shown in the lower left panel of Figure~\ref{fig:length}. In the D-web method, the adoption of larger thresholds can lead to a decrease in the number of short and long filaments throughout the range. For
$R_{\rm s} = 2.1 \mpc$, the total number of filaments $N_{\rm f} = 274~842, 178~417,  100~626$ and $48~719$ for $\lambda_{\rm th} = 0.1, 0.2, 0.4$ and $0.8$,
respectively.

In the lower right panel of Figure~\ref{fig:length}, the PDFs of different
thresholds are consistent with each other, indicating that the probability of filament
lengths does not change with the variation of threshold parameters.  PDF $f(L)$ with
fixed smoothing length $R_{\rm s} = 2.1 \mpc$ can be expressed by the best fit
relation $\log f(L) = -0.1 L -0.37$, shown by the black dashed line in the
lower right panel of Figure~\ref{fig:length}.

\subsection{Filament density profile}

\begin{figure}
\includegraphics[width=0.5\textwidth]{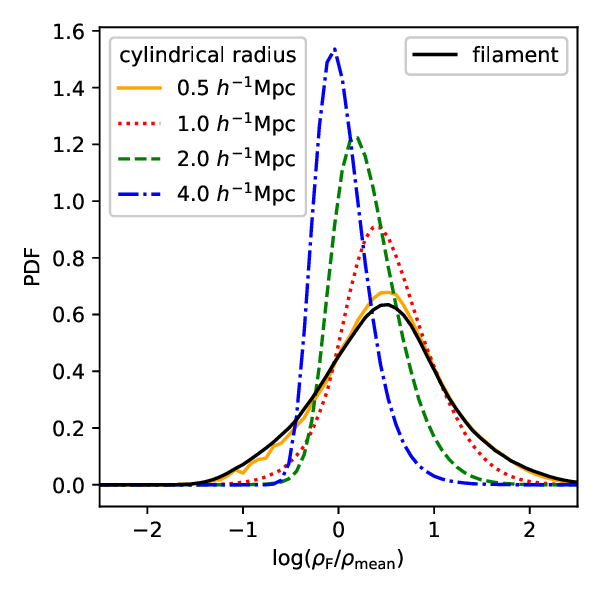}
\caption{Probability distribution function (PDF) of the logarithmic density contrast of the filament segments for varying the cylindrical radii. For a segment in a filament, the segment density is calculated by counting the number of dark matter particles in the  cylindrical segment with the radius of $r=0.5$ (orange), $1.0$ (red), $2.0$ (green), or $4.0 \mpc$ (blue). 
For reference, the black solid line illustrates the probability distribution function of the densities of filament cells 
determined by the COWS method.
}
\label{fig:pdf_radius}
\end{figure}

\begin{figure*}
\includegraphics[width=0.49\textwidth]{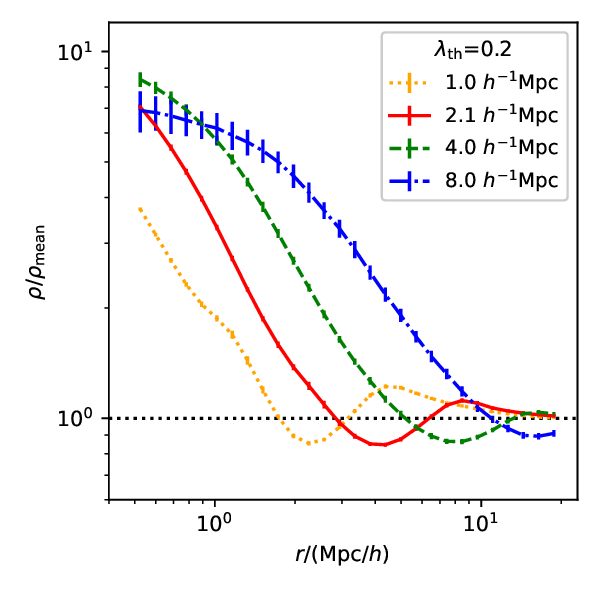}
\includegraphics[width=0.49\textwidth]{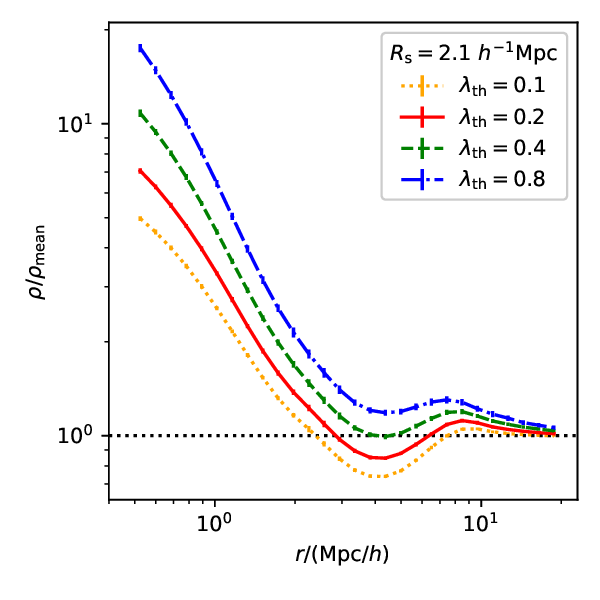}
\caption{Density profile of filaments as a function of the radial distance $r$ to
the filament axes. The left panel shows the radial density profile of filaments
detected by different values of smoothing lengths at a fixed threshold
$\lambda_{\rm th} = 0.2$, while the right panel shows the density profile of
filaments for different thresholds at a fixed smoothing length $R_{\rm s} = 2.1 \mpc$. 
The error bars are measured from $100$ bootstrap samples.} 
\label{fig:profile}
\end{figure*}

\begin{figure}
\includegraphics[width=0.5\textwidth]{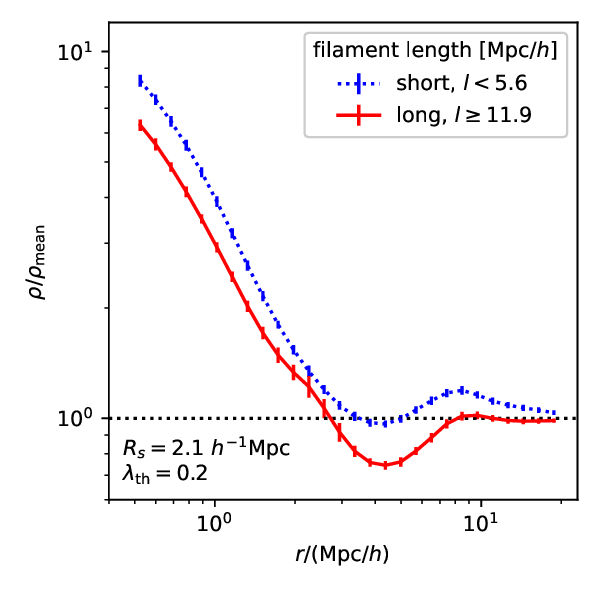}
\caption{Radial density profile of short (blue) and long (red) filaments  
based on the environment classification of $R_{\rm s} = 2.1 \mpc$ 
and $\lambda_{\rm th} = 0.2$. } 
\label{fig:profile2}
\end{figure}

\begin{figure}
\includegraphics[width=0.5\textwidth]{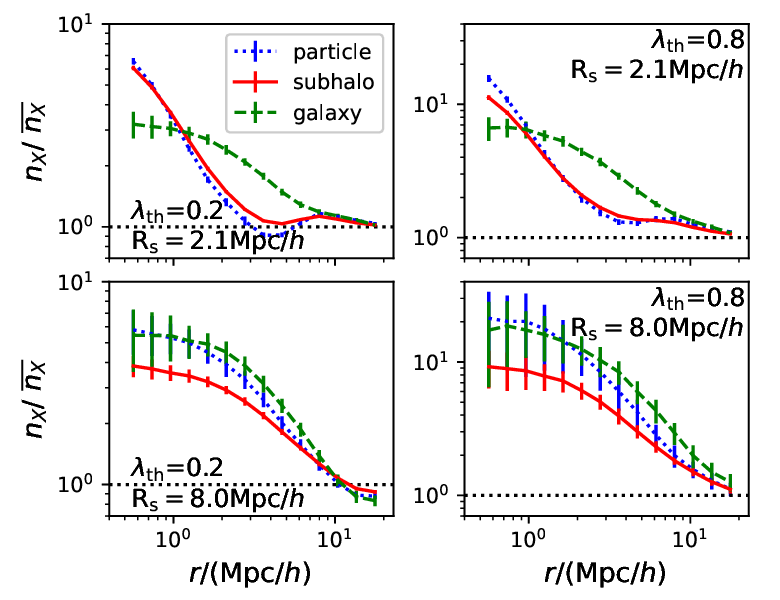}
\caption{Number density profiles of dark matter particles (blue), subhaloes (red), and galaxies (green)
around filaments.
The various panels display results for filaments categorised by different parameters $R_{\rm s}$ and
$\lambda_{\rm th}$. } 
\label{fig:number}
\end{figure}

In this section, we investigate the density distribution of filaments identified by
dark-matter particles in the ELUCID simulation. For each segment in a filament, 
the density of the segment is defined as 
\begin{equation}\label{eqn:density}
\rho_{\rm F} (\le r) = \frac{m_{\rm p}N_{\rm p}}{\pi r^2 l},
\end{equation}
where $m_{\rm p}$ is the particle mass,  $N_{\rm p}$ is the number of dark matter 
particles in the cylindrical segment,
$r$ is the radial distance to the axis of the segment, and $l$ represents the length of the segment, calculated as the distance between adjacent filament cells, expressed as $l=|\boldsymbol{p}_{i+1}-\boldsymbol{p}_i|$, where $\boldsymbol{p}_{i+1}$ and $\boldsymbol{p}_i$ are the positions of these neighbouring filament cells. According to Equation~\ref{eqn:density}, the mass density of the filament can be
calculated by counting the number of particles in the cylindrical segment within the
radius of $r$. 

Figure~\ref{fig:pdf_radius} shows the
probability distribution function of the logarithmic density contrast of
the filament segments for varying the cylindrical radii at $r=0.5, 1.0, 2.0,$ or $4.0 \mpc$, respectively. The COWS filament is made
up of a single pathway of connected cells, which is a subset of cells classified
as filaments in the density Hessian matrix method. For reference, the solid black line in Figure~\ref{fig:pdf_radius} depicts the density distribution of the filament cell determined by the COWS method. In particular, the filament cell density distribution (black) is very similar to the PDF for $r=0.5\mpc$ (orange), considering that the cell size in this investigation is around $0.5\mpc$. Clearly, the filament density is significantly influenced by the radius chosen for the cylindrical filament. Smaller radii result in higher filament densities and wider density distributions. The density of filament segments would rapidly approach the cosmic mean density ($\rho_{\rm mean}$) along with the increase of the cylindrical radius, leading to narrower distributions.

We subsequently evaluated the radial density profile of the filaments, which represents the density distribution along the orthogonal direction to the filament axis. A filament is composed of several interconnected segments. In practice, the radial density is assessed by investigating the dark-matter density distribution surrounding individual segments within the filaments. Each cylindrical segment $i$ is partitioned into cylindrical shells $40$ using uniform logarithmic bins centred on its axis, spanning from the core of the segment to $r=20\mpc$. The distribution of particles within each radial shell determines its density. Specifically, the density of the $j$-th shell of the segment $i$ is calculated as
\begin{equation}
\rho_{ij} = \frac {m_{\rm p} N_{ij}} {\pi (r_{j+1}^2-r_j^2) l_i}
\end{equation}
where the index $j$ denotes the $j$-th cylindrical shell with thickness $r_{j+1}-r_j$, $l_i$ is the length of segment $i$,
and $N_{ij}$ is the particle count within the $j$-th shell of segment $i$.
The average density $\rho_j$ for the $j$-th cylindrical shell is determined by computing the mean of the density profiles from all segments, as given by 
\begin{equation}
\rho_j = \frac{1}{N_s}\sum\limits_{i=1}^{N_s} \rho_{ij},
\end{equation}
where $N_s$ represents the total number of segments.

Figure~\ref{fig:profile} shows the density profile
of the filaments as a function of the radial distance $r$ from the filament axes. The
density $\rho$ has been normalised by the mean density $\rho_{\rm mean}$ of
the universe. 
The error bars depicted in Figure 9 are derived by bootstrapping the segment profiles utilising the method \texttt{astropy.stats.bootstrap} from the \texttt{Astropy} package\footnote{\href{http://www.astropy.org}{http://www.astropy.org}}. From the entire set of $N_t$ segments, a subset is chosen at random and their mean values are determined. This random selection and computation of averages is repeated 100 times, resulting in 100 different mean values. The standard error, derived from these 100 mean values, corresponds to the error bars presented in Figure 9.

The left panel of Figure~\ref{fig:profile} shows the radial density profile of the
filaments for different smoothing lengths $R_{\rm s} = 1.0, 2.1, 4.0$ and $8.0\mpc$ at a fixed threshold $\lambda_{\rm th} = 0.2$. Obviously, the density profile of the filaments depends on the smoothing scale $R_{\rm s}$. In general, the density
profile shown in Figure~\ref{fig:profile} agrees with the results of similar previous studies
\citep{Galarraga2020, Pfeifer2022, Galarraga2023}. The density profile of filaments exhibits an excess with respect to the mean density of the Universe.
For different smoothing lengths $R_{\rm s}$, the density profile decreases
with increasing radial distance $r$, until it reaches a valley approximately twice
the smoothing length, i.e., about $2R_{\rm s}$. It should be noted that the density profile that shows a valley and a minor peak in the outer regions of the filaments was also reported in previous studies by \citet{Pfeifer2022} and \citet{YangTianyi2022}. In contrast, in studies by \citet{Galarraga2020} and \citet{Galarraga2023}, these characteristics are less pronounced.
The valley in the radial density profile might stem from the technique used to generate the density field using CIC interpolation
in the COWS method. The COWS method depends on various types of cosmic web cells classified on a uniform grid, whereas DisPerSE detects filaments by analysing the topological features of the density field. This field can be obtained using either the Delaunay Tessellation Field Estimator (DTFE) or the CIC approach, based on the spatial distribution of the input particles. Using data from the MillenniumTNG simulation \citep{Hern2023}, we investigated the density profiles around filaments recognised by DisPerSE using both CIC and DTFE density constructions. We found that employing the CIC method in DisPerSE for generating density leads to a valley feature in the density profile, whereas this feature is absent when the DTFE method is used for density construction \citep{Galarraga2023,Wang2024}. 

The right side of Figure~\ref{fig:profile} illustrates the density profiles of filaments at varying thresholds $\lambda_{\rm th} = 0.1, 0.2, 0.4$ and $0.8$ with a constant smoothing length $R_{\rm s} = 2.1 \mpc$. Clearly, as the thresholds increase, so do the amplitudes of the density profiles, suggesting that larger thresholds capture filaments of higher density and greater significance. This observation aligns with Figure~\ref{fig:grid_cw}, where higher thresholds effectively exclude less dense, indistinct filaments.

Recently, \citet{Galarraga2020} claimed that short and long filaments have
different statistical properties. In this study, we investigate the density
profiles of short and long filaments for the catalogue of $R_{\rm s} = 2.1 \mpc$ and
$\lambda_{\rm th}=0.2$. We divide the filament catalogue into four different samples
so that each contains the same number of segments and the separation points are
$5.6$, $8.0$, and $11.9\mpc$. Figure~\ref{fig:profile2} shows the radial density profiles of the short ($L<5.6 \mpc$) and long ($L>11.9\mpc$) filaments for the cases of
$R_{\rm s} = 2.1 \mpc$ and $\lambda_{\rm th}=0.2$. We can see that the short and long
filaments have significantly different radial density profiles. Short filaments
have higher density profiles than long filaments at all radial distances, which
is in agreement with the filament density profiles by the DisPerSE method in
\citet{Galarraga2020}.

Next, we examine the number density profiles surrounding filaments for galaxies as well as subhaloes. 
The spatial distribution of galaxies in SDSS DR7 is associated with the dark-matter distribution in the constrained
ELUCID simulation. This linkage is due to the simulation's use of initial conditions constrained by the mass density field obtained from the SDSS DR7 galaxy distribution.

To ascertain the number density profiles of galaxies, the spatial coordinates (ra, dec, z) of galaxies recorded by SDSS are converted into the (X, Y, Z) coordinate system employed in the ELUCID simulation. Subsequently, we measured the number density of galaxies, subhaloes, and particles around the filament axes identified in the ELUCID simulation. To ensure a fair comparison, we excluded particles and subhaloes in the ELUCID simulation that fall outside the survey sky coverage area. Figure~\ref{fig:number} illustrates the number density profiles $n_X/\overline{n_X}$ from the filament axes, where $n_X$ is the number density
at the radial distance $r$ from the filament axes, $X$ represents galaxies, subhaloes or particles, and $\overline{n_X}$ signifies their mean value. The different panels in Figure~\ref{fig:number} present data for filaments classified by parameters $R_{\rm s} = 2.1, 8.0 \mpc$ and $\lambda_{\rm th} = 0.2, 0.8$. For particles shown by the blue dotted line in Figure~\ref{fig:number}, the number density profile corresponds to the overdensity $\rho/\rho_{\rm mean}$ displayed in Figure~\ref{fig:profile}, because $\rho/\rho_{\rm mean}$ is derived from particle distribution. The variation in number density profiles of galaxies and particles is influenced by the parameter $R_s$. When $R_s=2.1\mpc$, there is a notable difference between the number densities of galaxies and particles. However, at $R_s=8.0\mpc$, this difference decreases. In the case of subhaloes, their number density profile is similar to that of particles at $R_s=2.1\mpc$, but the divergence between subhaloes and particles becomes more pronounced at $R_s=8.0\mpc$.

\section{Summary}\label{sec_summary}

Using dark matter distribution from the cosmological and constrained ELUCID
simulation, we have investigated the statistical properties of filaments in
the cosmic web, such as the filament length distribution and the radial density
profile. The cosmic web type in space is classified by the Hessian matrix of
the density field. According to Equation~\ref{eqn:hessian}, the number of
eigenvalues above a certain threshold $\lambda_{\rm th}$ can be used to identify
knots, filaments, sheets, and voids based on smooth density with a Gaussian filter of the smoothing scale $R_{\rm s}$. 

To study the length of the filament and the profile of the radial density, the filament axes are identified by the medial axis thinning applied in the COWS method
\citep{Pfeifer2022}, based on the classifications of the cosmic web by the Hessian matrix of
the density field. As shown in Figure~\ref{fig:slice_filament}, the medial axes of the filament trace the spines of cosmic filaments very well characterized by
the distribution of dark-matter particles. Note that the direction of the filament
can also be indicated by the eigenvector $\hat{\boldsymbol{e}}_3$ of the density
Hessian matrix in Equation~\ref{eqn:hessian} \citep{Zhang2009, Zhang2013, Zhang2015}.
We confirm that the filament spines identified by the medial axis thinning
algorithm are strongly aligned with the eigenvector $\hat{\boldsymbol{e}}_3$ of
the density Hessian matrix, indicating the validity of the filament finding methods.

Although the medial axis thinning algorithm has no free parameter, the cosmic web
classification method has two parameters, the smoothing length $R_{\rm s}$ and
the threshold $\lambda_{\rm th}$, which can affect the statistical properties of
cosmic filaments. This research mainly examines how cosmic structures are influenced by the two parameters $R_{\rm s}$ and $\lambda_{\rm th}$.
The following is a summary of our findings.

\begin{enumerate}
\item [(i)]
The volume fraction of each type of cosmic web is strongly dependent on the
threshold parameter $\lambda_{\rm th}$. For knots, filaments, and sheets, volume
fractions decrease with increasing thresholds, while the volume fraction of voids
increases significantly with increasing $\lambda_{\rm th}$. In addition, the volume
fraction of each type of cosmic web is weakly dependent on the smoothing length
$R_{\rm s}$.

\item [(ii)] The filament length function varies with changes in the smoothing scales $R_{\rm s}$.
It reaches a maximum at $L \approx 2.5 R_{\rm s}$, from approximately $5.3 \mpc$ at $R_{\rm s} = 2.1 \mpc$ to approximately $20.0 \mpc$ at $R_{\rm s} = 8.0 \mpc$.

\item [(iii)]
In the entire range of filament length $L$, the filament length functions
decease with increasing thresholds $\lambda_{\rm th}$, demonstrating a decrease in the number of short and long filaments with increasing thresholds
$\lambda_{\rm th}$.

\item [(iv)]
The PDFs of the filament lengths are strongly dependent on the smoothing scales
$R_{\rm s}$ with $\log f(L) \propto L$, where $f(L)$ is the probability distribution
function. The PDFs of filament lengths are very similar for different thresholds. 
At $R_{\rm s} = 2.1 \mpc$, PDF can be expressed by the best-fitting relation
$\log f(L) = -0.1 L -0.37$. 

\item [(v)]
The mass density of the filament decreases with increasing radial distance to the filament axis. The density profile
of filaments exhibits an excess with respect to the mean density of the Universe.

\item [(vi)]
The radial density profile of the filaments depends on the parameters
$R_{\rm s}$ and $\lambda_{\rm th}$. It reaches a valley of approximately $2R_{\rm s}$. The amplitudes of the density profiles increase
with increasing thresholds $\lambda_{\rm th}$.

\item [(vii)]
The radial density profiles are different for short and long filaments. 
Short filaments have higher density profiles than long filaments in the whole
range of radial distance.
\end{enumerate}

In conclusion, the results presented in this paper show that the filament
length and the radial density profile depend on the smoothing scales
$R_{\rm s}$ and the thresholds $\lambda_{\rm th}$ in the filament-finding
method. Further analyses are useful in understanding the dependence of
filament properties on the adopted parameters using different filament-finding algorithms. Using this collection of filaments derived from the ELUCID dark matter distribution, future research will explore the extraction and quantification of these filaments with respect to galaxies, utilising both the ELUCID simulation and SDSS observations in an upcoming study.

\section*{Acknowledgements}

We express our gratitude to the anonymous referee for the insightful comments that greatly enhance the quality of this paper.
This work is supported by the National Natural Science Foundation of
China (Nos. 12273088, 11833005, 11890692, 11621303), the National Key 
R\&D Programme of China (2023YFA1607800, 2023YFA1607804), the CSST project
No. CMS-CSST-2021-A02, the Fundamental Research Funds for Central Universities, 111 project No. B20019 and the Shanghai Natural Science
Foundation, grant No.19ZR1466800. PW is sponsored by the Shanghai
Pujiang Programme (No. 22PJ1415100).

This work is also supported by the High-Performance Computing Resource in the
Core Facility for Advanced Research Computing at Shanghai Astronomical
Observatory.

This work used {\tt Astropy:} a community-developed core Python package
and an ecosystem of tools and resources for astronomy \citep{Astropy2013, Astropy2018, Astropy2022}. 

Funding for the Sloan Digital Sky Survey IV has been provided by the
Alfred P. Sloan Foundation, the U.S. Department of Energy Office of
Science, and the Participating Institutions. SDSS acknowledges support
and resources from the Centre for High-Performance Computing at the
University of Utah. The SDSS website is www.sdss.org.

SDSS is managed by the Astrophysical Research Consortium for the
Participating Institutions of the SDSS Collaboration including the
Brazilian Participation Group, the Carnegie Institution for Science,
Carnegie Mellon University, the Chilean Participation Group, the
French Participation Group, Harvard-Smithsonian Center for
Astrophysics, Instituto de Astrof{\'i}sica de Canarias, The Johns
Hopkins University, Kavli Institute for the Physics and Mathematics of
the Universe (IPMU)/University of Tokyo, Lawrence Berkeley National
Laboratory, Leibniz Institut f{\"u}r Astrophysik Potsdam (AIP),
Max-Planck-Institut f{\"u}r Astronomie (MPIA Heidelberg),
Max-Planck-Institut f{\"u}r Astrophysik (MPA Garching),
Max-Planck-Institut f{\"u}r Extraterrestrische Physik (MPE), National
Astronomical Observatories of China, New Mexico State University, New
York University, University of Notre Dame, Observat{\'o}rio Nacional/
MCTI, The Ohio State University, Pennsylvania State University,
Shanghai Astronomical Observatory, United Kingdom Participation Group,
Universidad Nacional Aut{\'o}noma de M{\'e}xico, University of
Arizona, University of Colorado Boulder, University of Oxford,
University of Portsmouth, University of Utah, University of Virginia,
University of Washington, University of Wisconsin, Vanderbilt
University, and Yale University.

\section*{Data availability}
The data underlying this article will be shared on reasonable request to the corresponding author.

\bibliographystyle{mnras}
\bibliography{bibliography}

\bsp    
\label{lastpage}
\end{document}